\documentclass[
superscriptaddress
,preprint
,tightenlines
,showpacs
,showkeys
,nofootinbib
,aps
,amsfonts
,amssymb
,byrevtex
,pdftex 
]{revtex4-1}

\pdfoutput=1
\usepackage{graphicx} 
\usepackage{amsmath, amssymb, bm} 
\usepackage[dvips]{color}

\usepackage{psfrag,epsfig}
\usepackage{natbib}
\usepackage{hyperref}
\usepackage{slashed}

\usepackage{braket}
\usepackage{float}
\usepackage{subcaption}
\usepackage{ragged2e}
\DeclareCaptionJustification{justified}{\justifying}
\usepackage{multirow}
\usepackage{dcolumn}
\usepackage{booktabs}

\usepackage{csvsimple} 

\usepackage[normalem]{ulem}  

\def\){\right)} 
\def\({\left(} 
\def\]{\right]} 
\def\[{\left[}

\def\affHISKP{\affiliation{Helmholtz-Institut f\"ur Strahlen- und Kernphysik~(Theorie) \\ 
and Bethe Center for Theoretical Physics, Universit\"at Bonn, D-53115 Bonn, Germany}}
\def\affKMU{\affiliation{Department of Physics, Karamanoglu Mehmet Bey University, Karaman 70100, Turkey}}

\begin{document}


\title{Lattice calculations for two-component fermion systems with unequal masses: one dimension}

\author{Serdar~Elhatisari}
\email{elhatisari@hiskp.uni-bonn.de}
\affHISKP
\affKMU

\begin{abstract}
We consider systems of two-component fermions with unequal masses and interacting via a short-range attractive potential. We discuss the case where the two-component fermions form a shallow dimer with large scattering length. The three-fermion and four-fermion systems with such properties are universal and charazteried by the two-fermion scattering length $a_{\text{ff}}$ and the ratio of the mass of spin-$\uparrow$ fermion to the mass of spin-$\downarrow$ fermion, $m_{\uparrow}/m_{\downarrow}$. In this study using lattice effective field theory we analyze fermion-dimer and dimer-dimer systems, and calculate the universal fermion-dimer and dimer-dimer scattering lengths for various values of the mass ratio $m_{\uparrow}/m_{\downarrow}$. We find that these universal scattering lengths increase logarithmically with the mass ratio $m_{\uparrow}/m_{\downarrow}$.

\end{abstract}

\date{\today}
\maketitle


\section{\textbf{Introduction}}
\label{sec:intr}

Low-energy universality is an important phenomenon in several branches of physics and it appears at large scattering length where the physics is insensitive to the details of the short-range potentials~\cite{Braaten:2004rn}. This allows us to connect the physics at different scales in an elegant way. In nuclear physics, the nucleon-nucleon scattering lengths are much larger than all other length scales, and in very low-energy limit the systems can be described by only local contact interactions~\cite{vanKolck:1998bw}. In atomic physics, the van der Waals interactions between alkali atoms can be approximated by short-range interactions at sufficiently low-energies. Also in the physics of ultracold atoms, this phenomenon can be realized by tuning the scattering length arbitrarily near a Feshbach resonance using an external magnetic field as a tool~\cite{Feshbach:1962ut,Chin:2010}. 

The theoretical studies of low-energy universality go beyond two-body systems. A diagrammatic approach has been developed and used to study 3-body and 4-body systems consisting of two-component fermions~\cite{Levinsen:2010mn}. Lattice effective field theory has been used to extract the universal scattering length and effective range in the fermion-dimer and the dimer-dimer system~\cite{Elhatisari:2016hui}.

Progress in a few-body problems is crucial since the knowledge of the scattering properties of composed systems is of significant importance in understanding the dynamics of the many-body systems. In this paper we study  systems of two-component fermions forming a shallow dimer with large scattering length. For the sake of simplicity our analysis concentrates on a few-body problem in one spatial dimension. In our analysis the masses of different particle species  are not necessarily equal to each other.

For the system of particles interacting via a finite-range potential, at low energies the scattering phase shifts $\delta(p)$ is parameterized by the effective range expansion,
\begin{align}
p \ \tan\delta(p) = \frac{1}{a_{\rm ff}} + \frac{1}{2} \ r_{\rm ff} \ p^2 + \ldots \,,
\label{eqn:ERE-001}
\end{align}
where $p$ is the relative momentum between two fermions, $a_{\rm ff}$ is the scattering length, and $r_{\rm ff}$ is the effective range. In the zero-range limit, the scattering length is related to the dimer binding energy by the formula
\begin{align}
B_{\rm d} = 1/(2\mu \ a_{\rm ff}^2) \,,
\label{eqn:BE-001}
\end{align}
where $\mu$ is the reduced mass.

\section{\textbf{Lattice formalism}}
\label{sec:latticeformalism}

In this Section, following Refs.~\cite{Lee:2008fa,Elhatisari:2016hui} we introduce the lattice theory of two-component fermions. We work with natural units where $\hbar = c = 1$, and we denote the two components as spin-$\uparrow$ and spin-$\downarrow$ with masses $m_{\uparrow}$ and $m_{\downarrow}$, respectively. The non-relativistic Hamiltonian in the continuum is, 
\begin{align}
\hat{H} = &
\sum_{s} \frac{1}{2 m_{s}} \, 
\int  \, dr \,
\nabla b^{\dagger}_{s}(r) \,
\nabla b^{\,}_{s}(r) +
C_0 \,
\int  \, dr \,
b^{\dagger}_{\uparrow}(r) \, b^{\,}_{\uparrow}(r) \, \,
b^{\dagger}_{\downarrow}(r) \, b^{\,}_{\downarrow}(r)
\,,
\end{align}
where $s$ labels the particle species, $C_0$ is the zero-range interaction strength, and $b^{\,}_{s}$ and $b^{\dagger}_{s}$ are the annihilation and creation operators, respectively.

In our calculations, we utilize a lattice that is periodic with the lattice spacing $a$, and we define all physical quantities in lattice units (l.u.) multiplying them by the corresponding powers of $a$. Therefore, the non-relativistic lattice Hamiltonian with $\mathcal{O}(a^{4})$-improved action~\cite{Lee:2008fa} is, 
\begin{align}
H = &
\sum_{s} \frac{1}{2 m_{s}} \, 
\sum_{n}  \,
\left[
\sum_{k = -3}^{3}
w_{|k|}
b^{\dagger}_{s}(n) \,
b^{\,}_{s}(n+k)
\right]
+
C_0 \,
\sum_{n}  \,
b^{\dagger}_{\uparrow}(n) \, b^{\,}_{\uparrow}(n) \, \,
b^{\dagger}_{\downarrow}(n) \, b^{\,}_{\downarrow}(n)
\,,
\label{eqn:Hamiltonian-001}
\end{align}
where $n$ labels the lattice sites, and $w_0$, $w_1$, $w_2$, $w_3$ are the hopping parameters and their values are $49/18$, $-3/2$, $3/20$, $-1/90$, respectively.

In the calculations, the interaction strength $C_0$ is tuned to produce a two-body scattering length much larger than the potential range. When the scattering length is positive and large, there exists a shallow bound dimer given by Eq.~(\ref{eqn:BE-001}).

For convenience we set parameters to values for systems of nuclear physics. However, we present the final results in terms of the two-body scattering length $a_{\rm ff}$, which are completely independent of chosen values and scales. We choose the masses $m_{\uparrow} = 1$~GeV and change the value of $m_{\downarrow}$ such that we can analyze systems in the limits $m_{\uparrow} \to \infty$ and $m_{\downarrow} \to \infty$. Therefore, in these limits, the reduced mass $\mu$ is kept constant and it equals to the mass of the light particle. Also, in order to remove any discretization error in the final results we repeat our calculation for various values of the two-body scattering length, $a_{\rm ff}$, ranging from $1.4$~fm to $10$~fm which are used to perform the continuum limit extrapolations.

\section{\textbf{Scattering on the lattice}}
\label{sec:scatteringonthelattice}

The direct information that can be obtained from the lattice calculations is the energy levels. However, L\"uscher found an elegant relation of the two-body energy levels for a \emph{periodic} lattice with the elastic scattering phase shifts in the infinite volume and continuum limits~\cite{Luscher:1986pf,Luscher:1990ux}. The scattering information can also be extracted by using the wave functions on the lattice as well as energy levels~\cite{Borasoy:2007vy,Rokash:2015hra,Lu:2015riz}. In this work, we use the L\"{u}scher method, and in the following we briefly discuss it for one dimension case.

Let us consider a two-body system with zero total momentum and a potential of a finite-range $R$ on a periodic lattice of size $L$. Then the wave function at distances $r > R$ takes the asymptotic form $\psi(r) \sim \cos[p r +\delta(p)]$, and due to the periodicity it satisfies the condition $\psi(La/2) = \psi(-La/2)$ and $\partial_{r}\psi(r) |_{La/2} = \partial_{r}\psi(r) |_{-La/2}$, which yields 
\begin{align}
pLa + 2\delta(p) = 2n\pi 
\,,
\quad 
n = 0,1,2,\dots
\label{eqn:Luescherformula}
\end{align}
This relation gives us direct access to the scattering information using the lattice data for $p$ and the lattice parameter $L$.

\begin{figure}[!ht]
	\centering
	\includegraphics[width=0.7\linewidth]{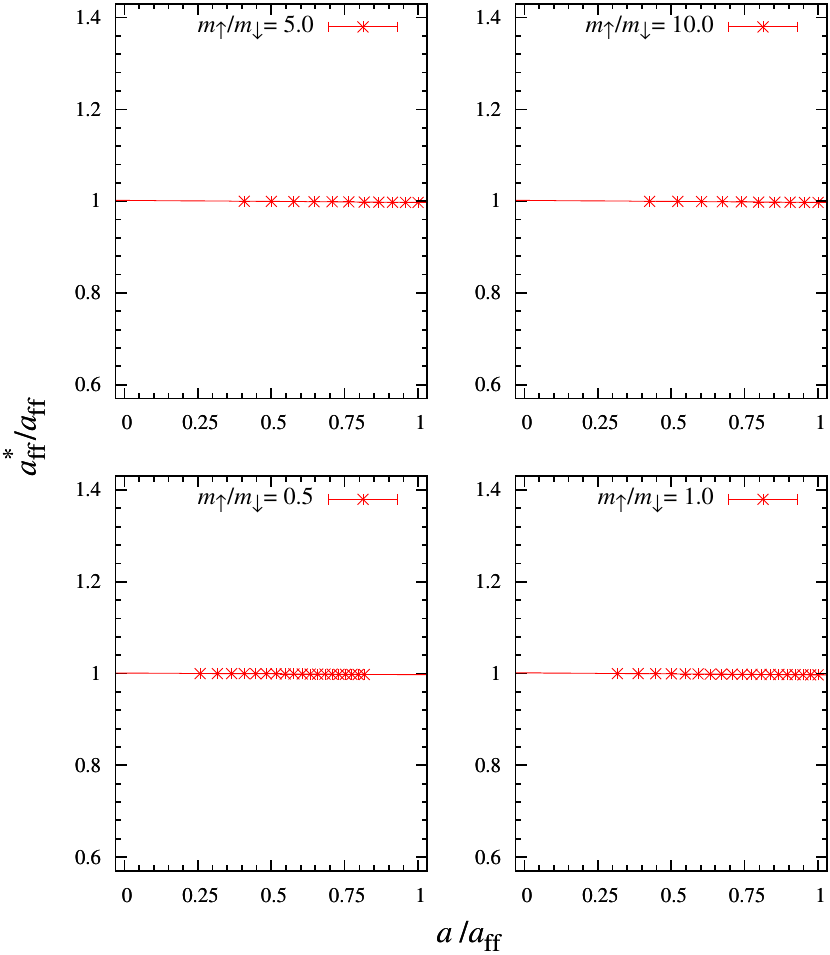}
	\caption{The ratio of the two-fermion scattering length $a_{\rm ff}^{*}$ extracted from the lattice data using L\"uscher’s formula to $a_{\rm ff}$ calculated from Eq.~(\ref{eqn:BE-001}). The results are plotted versus the lattice spacing $a$ as a fraction of  $a_{\rm ff}$, and fitted to a linear function of  $a/a_{\rm ff}$.}
	\label{fig:a_ff}
\end{figure}

We first perform the calculations for the two-fermion system. We tune the interaction strength $C_0$ to produce the dimer binding energy $B_{\rm d}$ in the infinite volume limit so that the finite volume effects are eliminated. Then we compute the low-energy spectrum of the two-fermion
lattice Hamiltonian at different values of $L$. The relative momentum $p$ to be used as input in Eq.~(\ref{eqn:Luescherformula}) is calculated from these energy levels by $p = \sqrt{2\mu E}$. In Fig.~\ref{fig:a_ff} we show the two-fermion scattering lengths from the lattice calculations for a few different mass rations $m_{\uparrow}/m_{\downarrow}$ and for various lattice spacing $a$.
 Here $a_{\rm ff}^{*}$ is extracted from the effective range expansion Eq.~(\ref{eqn:ERE-001}) by using the scattering phase shifts $\delta(p)$ and the relative momentum $p$ from the lattice calculations. The scattering length $a_{\rm ff}^{\,}$ is calculated by Eq.~(\ref{eqn:BE-001}) using the binding energy, $B_{\rm d}$ as input. The results from Fig.~\ref{fig:a_ff} show that we extract the scattering length from the lattice calculations with negligible lattice artifacts.
\section{\textbf{Results and Discussion}}
\label{sec:results-and-discussion}

\subsection{Fermion-dimer scattering}
\label{sec:fermion-dimer}

Now we discuss the calculation for the fermion-dimer system consisting of two spin-$\uparrow$ and one spin-$\downarrow$ fermions. In our fermion-dimer system we only consider the two-body interaction since the interactions beyond two-body interaction are irrelevant operators in the low energy physics of the two-component fermions~\cite{Elhatisari:2016hui}. 

We use the Lanczos  eigenvector method~\cite{Lanczos} to compute the low-energy spectrum of the lattice Hamiltonian at different values of $L$. Then we employ the L\"uscher method to extract the scattering phase shifts from the lattice data. To achieve this we need to compute the fermion-dimer relative momentum from the low-energy spectrum correctly, which is not straightforward as it is in the case of two-point like particles. First, at nonzero lattice spacing the effective mass of the dimer is not equal to $m_{\uparrow}+m_{\downarrow}$. Therefore, we compute the dimer effective mass by fitting Eq.~(\ref{eqn:disp-rel-001}) to the lattice dispersion relation of the dimer.
\begin{align}
D(p,m_{\rm d})= c_{0} \, \frac{p^2}{2\, m_{\rm d}}
+ c_{1} \, p^4 + \ldots \,,
\label{eqn:disp-rel-001}
\end{align}
where $c_{i}$ are the coefficients to be determined by the fit, $p$ is the total momentum of the moving dimer, and $m_{\rm d}$ is the physical dimer mass.

Secondly, in the fermion-dimer system, the existence of a moving dimer induces phase-twisted boundary conditions on the dimer’s relative-coordinate wave function. This effect results in a topological energy correction~\cite{Bour:2011ef}. When this effect is taken into account, the relative momentum of the fermion-dimer is determined by,
\begin{align}
E_{\rm fd}^{L} = \frac{p^2}{2\mu_{\rm fd}^{*}}
-B_{\rm d} 
-\Delta B_{d}^{L} \,  \cos(p \ a  \ L \ \alpha)
\label{eqn:fd-rel-mom-001}
\end{align}   
where $E_{\rm fd}^{L}$ is the fermion-dimer energy at lattice size $L$, $\Delta B_{d}$ is the finite volume correction of the dimer binding energy $\Delta B_{d} = B_{d}^{L}-B_{d}$, $\alpha = m_{\uparrow}/(m_{\uparrow} + m_{\downarrow})$, and $\mu_{\rm fd}^{*}$ is the fermion-dimer reduced mass calculated using the lattice-determined dimer effective mass. We solve Eq.~(\ref{eqn:fd-rel-mom-001}) for the relative
 momentum $p$ corresponding to lattice size $L$ using lattice energies $E_{\rm fd}^{L}$, and $B_{d}^{L}$ as input.

Since the one dimensional problem is integrable and exactly solvable using the Bethe Ansatz~\cite{Bethe1931}, therefore, we compare the lattice results with the Bethe Ansatz calculations.
\begin{figure}[!ht]
	\centering
	\includegraphics[width=1.\linewidth]{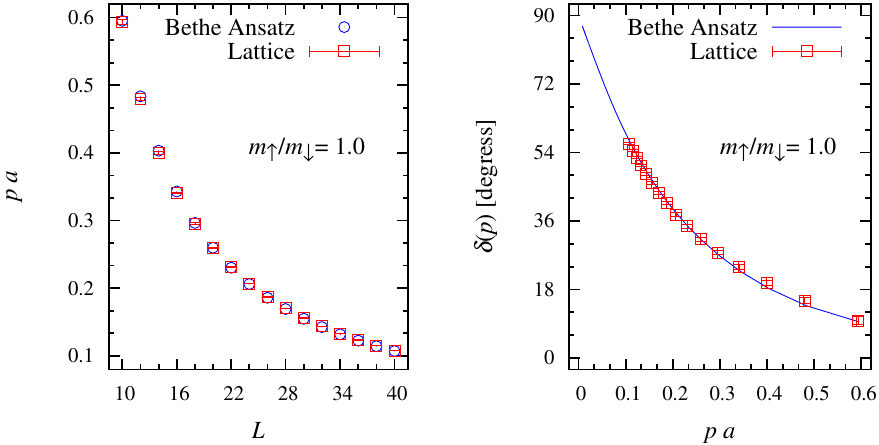}
	\caption{(Left) The fermion-dimer relative momentum $p\ a$ versus lattice size $L$, and (right) the scattering phase shifts $\delta(p)$ versus the relative momentum $p \ a$. The open squares are the results from the lattice calculations, and the open circles are the Bethe Ansatz results.}
	\label{fig:lattice-vs-BA}
\end{figure}
The results in Fig.~\ref{fig:lattice-vs-BA} clearly show that the effect of the topological correction due to the composite system and effect on the dimer effective mass due to a nonzero lattice spacing are removed from the relative momentum. The error on the lattice results in Fig.~\ref{fig:lattice-vs-BA} are the propagated error from one standard deviation of the error on the dimer effective mass due to the fit to the lattice dispersion relation of the dimer in Eq.~(\ref{eqn:disp-rel-001}).

\begin{figure}[!ht]
	\centering
	\includegraphics[width=0.5\linewidth]{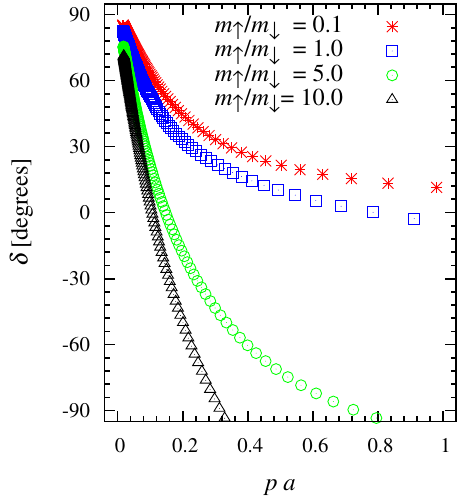}
	\caption{The scattering phase shift versus the relative momentum between fermion and dimer. In these calculations we set $a_{\rm ff}/a=5$.}
	\label{fig:deltaofp-fermidimer}
\end{figure}

The results for the fermion-dimer phase shifts are shown in Fig.~\ref{fig:deltaofp-fermidimer}. We plot the fermion-dimer scattering phase shifts as a function of the relative momentum between fermion and dimer for various values of the mass ratio $m_{\uparrow}/m_{\downarrow}$.
In each calculation we make a fit using the phase shifts and the relative momentum in the truncated effective range expansion,
\begin{align}
a_{\rm ff} \, p \ \tan\delta(p) =
\frac{1}{a_{\rm fd}/a_{\rm ff}} + \frac{1}{2} \ \left(r_{\rm fd}/a_{\rm ff}\right) \ \left(a_{\rm ff} p\right)^2 + \ldots \,,
\label{eqn:ERE-002}
\end{align}
where $a_{\rm fd}$ and $r_{\rm fd}$ are the fermion-dimer scattering length and effective range, respectively. We perform these calculations for various values of the two-body scattering length $a_{\rm ff}$, and we extract $a_{\rm fd}$ using Eq.~(\ref{eqn:ERE-002}). Then,  in order to remove the lattice discretization errors, we perform continuum limit extrapolations for these lattice results by making linear fits.
\begin{figure}[!ht]
	\centering
	\includegraphics[width=0.5\linewidth]{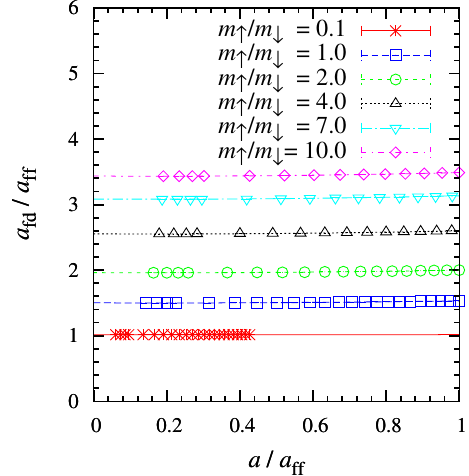}
	\caption{Plots of the scattering length extrapolations to the limit $a/a_{\rm ff} \to 0$ for various values of the mass ratio $m_{\uparrow}/m_{\downarrow}$.}
	\label{fig:afdoveraff}
\end{figure}

The results are shown in Fig.~\ref{fig:afdoveraff}. We plot the continuum limit extrapolation of the fermion-dimer scattering length $a_{\rm fd}$ as a fraction of the fermion-fermion scattering length $a_{\rm ff}$. This ratio $a_{\rm fd}/a_{\rm ff}$ is universal, and it is called the universal fermion-dimer scattering length. As it can be seen, the lattice discretization errors are negligible for smaller mass ratio $m_{\uparrow}/m_{\downarrow}$, while the continuum limit extrapolation is necessary as $m_{\uparrow}\to \infty$.

\begin{figure}[!ht]
	\centering
	\includegraphics[width=0.5\linewidth]{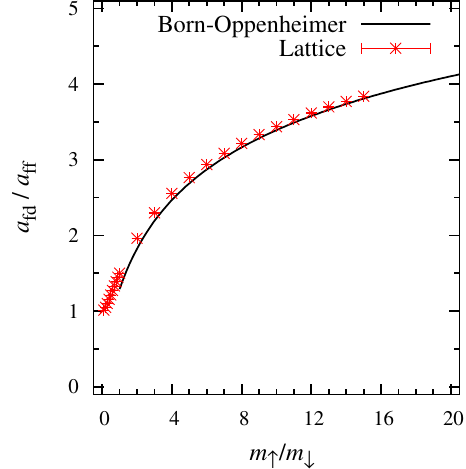}
	\caption{Plot of the universal fermion-dimer scattering phase shifts versus the mass ratio $m_{\uparrow}/m_{\downarrow}$. The asterisk points are the lattice result, and the solid line is the Born-Oppenheimer result.}
	\label{fig:afdoveraff_vs_Mm}
\end{figure}
In Fig.~\ref{fig:afdoveraff_vs_Mm} we plot the continuum limit extrapolated lattice results of the universal fermion-dimer scattering length $a_{\rm fd}/a_{\rm ff}$ versus the mass ratio $m_{\uparrow}/m_{\downarrow}$.

For the case of $m_{\downarrow} \to \infty$, the spin-$\downarrow$ particle is barely diffusing in space, and it can be regarded as a stationary particle. Furthermore, without loss of generality we can take its position to be at the origin, then we have a static attractive delta-function potential at the origin for the two spin-${\uparrow}$ particles. In this system, one of the spin-${\uparrow}$ particles is part of the dimer, and it already occupies the bound-state wave function of the attractive delta-function. This bound-state wave function is exactly orthogonal to all the scattering states of the delta-function potential. Therefore, the second spin-${\uparrow}$ 
particle scatters off the delta-function potential without caring at all about the other spin-${\uparrow}$ particle bound to the delta-function, and as $m_{\downarrow} \to \infty$ the fermion-dimer scattering length is approaching to the two-particle scattering length, $a_{\rm fd}/a_{\rm ff} \to 1$.

For the case of $m_{\uparrow} \to \infty$, the light spin-$\downarrow$ is exchanged between the two heavy spin-$\uparrow$ particles, and this induces an effective potential with a range proportional to $m_{\downarrow}^{-1}$. As a result of this interaction, the universal fermion-dimer scattering length increases with the mass ratio  $m_{\uparrow}/m_{\downarrow}$.

In the limit $m_{\uparrow} \to \infty$, the fermion-dimer system can be solved using the Born-Oppenheimer approximation. We also study the Born-Oppenheimer method for the fermion-dimer system to benchmark our lattice results, see Sec.~\ref{sec:BO-fermiondimer}. As discussed in Sec.~\ref{sec:BO-fermiondimer} the Born-Oppenheimer potential between the dimer and fermion is purely repulsive, and this results in increasing universal fermion-dimer scattering length with increasing mass ratio. In Fig.~\ref{fig:afdoveraff_vs_Mm} we show the results for the universal fermion-dimer scattering length $a_{\rm fd}/a_{\rm ff}$ calculated by solving Eq.~(\ref{eqn:H-BO-005}) numerically. We find a very good agreement between these two different methods in the limits as $m_{\uparrow} \to \infty$, where the Born-Oppenheimer approximation works.

\subsection{Dimer-dimer scattering}
\label{sec:dimer-dimer}
In this section we discuss the dimer-dimer system consist of two spin-$\uparrow$ and two-$\downarrow$ fermions. Here we consider only the two-body interaction since the higher-body interactions are irrelevant in the low energy limit~\cite{Elhatisari:2016hui}. We use the lattice Hamiltonian given in Eq.~(\ref{eqn:Hamiltonian-001}) at different values of $L$ and compute the dimer-dimer low energy spectrum using the Lanczos eigenvector method~\cite{Lanczos}.

\begin{figure}[!ht]
	\centering
	\includegraphics[width=0.5\linewidth]{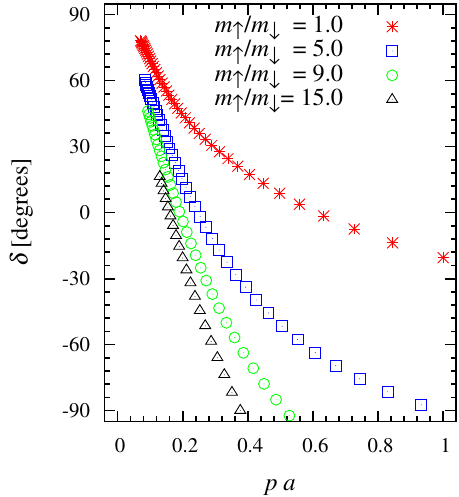}
	\caption{The scattering phase shift versus the relative momentum between dimers.  In these calculations we set $a_{\rm ff}/a=5$.}
	\label{fig:deltaofp-dimerdimer}
\end{figure}

The dimer-dimer scattering phase shifts are extracted using the L\"uscher method, and the results are shown in Fig.~\ref{fig:deltaofp-dimerdimer}. We plot the dimer-dimer scattering phase shifts as a function of the relative momentum between dimers for various values of the mass ratio $m_{\uparrow}/m_{\downarrow}$.

The relative momentum between dimers is computed in a similar manner as in the fermion-dimer case in Sec.~\ref{sec:fermion-dimer}. We use the dimer effective mass computed by Eq.~(\ref{eqn:disp-rel-001}). In the dimer-dimer system since  we have two bound dimers, we need to take into account the topological phase correction in the relative momentum calculation. Therefore, the relative momentum between two dimers are given by
\begin{align}
E_{\rm dd}^{L} = \frac{p^2}{2\mu_{\rm dd}^{*}}
-2 \ B_{\rm d} 
-2 \ \Delta B_{d}^{L} \,  \cos(p \ a  \ L \ \alpha)
\label{eqn:dd-rel-mom-001}
\end{align}   
where $E_{\rm dd}^{L}$ is the dimer-dimer energy at lattice size $L$, and $\mu_{\rm dd}^{*}$ is the dimer-dimer reduced mass calculated using the lattice-determined dimer effective mass. We solve Eq.~(\ref{eqn:dd-rel-mom-001}) for the relative momentum $p$ corresponding to lattice size $L$ using lattice energies $E_{\rm dd}^{L}$, and $B_{d}^{L}$ as input.

The computed scattering phase shifts using the lattice data in the L\"uscher method are used in the following truncated effective range expansion to extract the scattering length,
\begin{align}
a_{\rm ff} \, p \ \tan\delta(p) =
\frac{1}{a_{\rm dd}/a_{\rm ff}} + \frac{1}{2} \ \left(r_{\rm dd}/a_{\rm ff}\right) \ \left(a_{\rm ff} p\right)^2 + \ldots \,,
\label{eqn:ERE-005}
\end{align}
where $a_{\rm dd}$ is the dimer-dimer scattering length, and $r_{\rm dd}$ is the dimer-dimer effective range. The scattering length $a_{\rm dd}$ is computed for various values of the two-body scattering length $a_{\rm ff}$, then continuum limit extrapolations are performed for these lattice results by making linear fits in order to remove any lattice discretization error. 
\begin{figure}[!ht]
	\centering
	\includegraphics[width=0.5\linewidth]{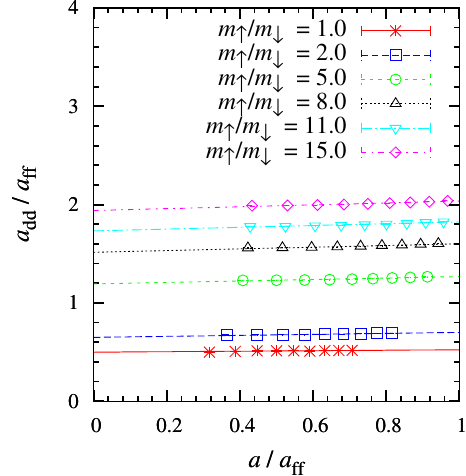}
	\caption{Plots of the scattering length extrapolations to the limit $a/a_{\rm ff} \to 0$ for various values of the mass ratio $m_{\uparrow}/m_{\downarrow}$.}
	\label{fig:addoveraff}
\end{figure}
The results are shown in Fig.~\ref{fig:addoveraff} where we plot the continuum limit extrapolation of the dimer-dimer scattering length $a_{\rm dd}$ as a fraction of the fermion-fermion scattering length $a_{\rm ff}$. This ratio $a_{\rm dd}/a_{\rm ff}$ is universal, and it is called the universal dimer-dimer scattering length.

\begin{figure}[!ht]
	\centering
	\includegraphics[width=0.5\linewidth]{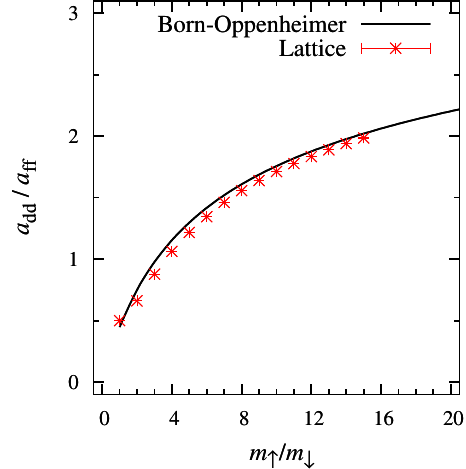}
	\caption{Plot of the universal dimer-dimer scattering phase shifts versus the mass ratio $m_{\uparrow}/m_{\downarrow}$. The asterisk points are the lattice result, and the solid line is the Born-Oppenheimer result.}
	\label{fig:addoveraff_vs_Mm}
\end{figure}
In Fig.~\ref{fig:addoveraff_vs_Mm} we plot the continuum limit extrapolated results of the universal dimer-dimer scattering length $a_{\rm dd}/a_{\rm ff}$ versus the mass ratio $m_{\uparrow}/m_{\downarrow}$.

In the limit $m_{\uparrow} \to \infty$, the spin-$\downarrow$ fermions are exchanged between the two heavy spin-$\uparrow$ particles, and this induces an effective potential with a range proportional to $m_{\downarrow}^{-1}$. As a result of this interaction, the universal dimer-dimer scattering length increases with the mass ratio  $m_{\uparrow}/m_{\downarrow}$.

We also study the Born-Oppenheimer method for the dimer-dimer system to benchmark our lattice results in the limit $m_{\uparrow} \to \infty$, see Sec.~\ref{sec:BO-dimerdimer}. As discussed in Sec.~\ref{sec:BO-dimerdimer} the Born-Oppenheimer potential between the dimers is repulsive except at very short distances, and when compared to the fermion-dimer system, the Born-Oppenheimer potential between the dimers is rather complicated. The reader can find a detailed discussion on the effective potential between the dimers in Ref.~\cite{Rokash:2016tqh}. Therefore, this repulsive Born-Oppenheimer potential between the dimers results in increasing universal dimer-dimer scattering length with increasing mass ratio. In Fig.~\ref{fig:addoveraff_vs_Mm} we show the results for the universal dimer-dimer scattering length $a_{\rm dd}/a_{\rm ff}$ calculated by solving Eq.~(\ref{eqn:H-BO-106}) numerically. We find a very good agreement between these two different methods in the limits as $m_{\uparrow} \to \infty$, where the Born-Oppenheimer approximation works. 

\section{\textbf{Conclusion}}
\label{sec:d}

In the low-energy limit or at large particle separation the systems consist of the two-component fermions have universal properties, and all properties of these systems scale proportionally with the fermion-fermion scattering length $a_{\rm ff}$.

In this study we have used lattice effective field theory and considered two-component fermions with different masses interacting via short-range interactions. We have computed the fermion-dimer scattering length $a_{\rm fd}$ and dimer-dimer scattering length $a_{\rm dd}$ in the universal limit of large fermion-fermion scattering length $a_{\rm ff}$. We have repeated our calculations using various values of the two-body scattering length $a_{\rm ff}$, then we have performed the continuum limit extrapolations of the lattice results to remove the lattice discretization errors. 

In the case of the two-component fermions with different masses, the mass ratio is a new parameter and it changes some of the properties of the system. Therefore, we have presented our final results of the universal fermion-dimer scattering length and universal dimer-dimer scattering length for various mass ratios in Figs.~\ref{fig:afdoveraff} and \ref{fig:addoveraff}. We have found that the universal fermion-dimer scattering length increases logarithmically with the mass ratio $m_{\uparrow}/m_{\downarrow}$ as shown in Figs.~\ref{fig:afdoveraff_vs_Mm} and \ref{fig:addoveraff_vs_Mm}.

Knowledge of the scattering properties of composed systems is of significant importance in understanding the dynamics of the many-body systems. For this purpose we have analyzed the scattering properties of the fermion-dimer and dimer-dimer systems and the mass ratio dependence of the universal fermion-dimer and universal dimer-dimer scattering lengths.

\section*{\textbf{Acknowledgment}} \label{sec:e}
The author is grateful to Dean~Lee and Ulf-G.~Mei{\ss}ner  for useful discussions and carefully reading the manuscript. This work is supported by the Scientific and Technological Research Council of
Turkey (TUBITAK) project no. 116F400, and the DFG (TRR 110).


\appendix
\section{The Born-Oppenheimer approach}

\subsection{Fermion-dimer}
 \label{sec:BO-fermiondimer}

Let consider two spin-${\uparrow}$ and one spin-${\downarrow}$ particles interacting via delta-function potential. The Hamiltonian of the systems is,
\begin{align}
H_{\rm fd} = 
-\frac{1}{2 m_{1}} \, \partial_{x_{1}}^{2}
-\frac{1}{2 m_{2}} \, \partial_{x_{2}}^{2} 
-\frac{1}{2 m_{3}} \, \partial_{x_{3}}^{2} 
+c_{0} \left[ \delta(x_{3}-x_{1}) + \delta(x_{3}-x_{2})\right]
\,,
\label{eqn:H-BO-001}
\end{align}
where $\partial_{x}^{2}=\partial^{2}/\partial x^{2}$ $m_{1}$, $m_{2}$, and $m_{3}$ are the masses and $x_{1}$, $x_{2}$, and $x_{3}$ are the coordinates of the spin-${\uparrow}$, spin-${\uparrow}$, and spin-${\downarrow}$ particles, respectively. Eq.(\ref{eqn:H-BO-001}) can be rewritten as
\begin{align}
H_{\rm fd} = 
-\frac{1}{2 \mu_{2}} \, \partial_{x}^{2}
-\frac{1}{2 \mu_{3}} \, \partial_{y}^{2} 
+c_{0} \left[ \delta(y-x/2) + \delta(y+x/2))\right]
\,,
\label{eqn:H-BO-002}
\end{align}
where
\begin{align}
&m_{1} = m_{2} = m_{\uparrow}\,,\nonumber\\
&m_{3} = m_{\downarrow}\,,\nonumber\\
&\mu_{2} = m_{\uparrow}/2\,,\nonumber\\
&\mu_{3} = 2m_{\uparrow}m_{\downarrow}/(2m_{\uparrow}+m_{\downarrow})\,,\nonumber\\
&x = x_{2}-x_{1}\,,\nonumber\\
&y = \frac{m_{1} x_{1}+m_{2} x_{2}}{m_{1}+m_{2}} -x_{3}\,.
\end{align}
The Schr\"odinger equation in the limit $m_{\uparrow} \to \infty$ is,
\begin{align}
-\frac{1}{2 \mu_{3}}\partial_{y}^{2} \phi(y;x)
+c_{0} [ \delta(y-x/2) 
+ \delta(y+x/2)]
\phi(y;x) 
= u(x)\phi(y;x)
\,.
\label{eqn:H-BO-003}
\end{align}
where $\phi(y;x)$ is the solution of Eq.~(\ref{eqn:H-BO-003}) for a fixed value of $x$. Similarly, the energy of the system, $u(x)$, is obtained for a fixed value of $x$.  Using boundary conditions, the continuity of the wave functions and the discontinuity of their first derivative at $y = \pm x/2$, we obtain the energy as a function of $x$,
\begin{align}
u_{\ell}(x)
=\frac{1}{2 \mu_{3}} 
\left[
-\beta
+
\frac{1}{x}
W((-1)^{\ell+1} x \, \beta\, e^{x \, \beta})
\right]^{2}
\,.
\label{eqn:H-BO-004}
\end{align}
where $\beta = c_{0}\mu_{3}$, $W(r)$ is the Lambert $W$ function, and $\ell = 0 \, (\ell = 1)$ gives the even (odd) solution.

Now, using $u_{\ell}(x)$, the solutions of  Eq.~(\ref{eqn:H-BO-003}), in Eq.~(\ref{eqn:H-BO-002}), and we can solve Eq.~(\ref{eqn:H-BO-002}). 
\begin{align}
-\frac{1}{2 \mu_{2}} \, \partial_{x}^{2}\psi(x)
+u_{\ell}(x)\psi(x)
=
E\psi(x)
\,,
\label{eqn:H-BO-005}
\end{align}
The total wave function of the fermion-dimer system, Eq.~(\ref{eqn:H-BO-002}), $\Psi(x,y) = \psi(x)\phi(y;x)$, is antisymmetric under exchange of $x_{1} \leftrightarrow x_{2}$. Therefore, the Born-Oppenheimer solution of Eq.~(\ref{eqn:H-BO-001}) is obtained by solving the Schr\"odinger equation for $\ell = 1$.

\subsection{Dimer-dimer}
\label{sec:BO-dimerdimer}

Now we consider two spin-${\uparrow}$ and two spin-${\downarrow}$ particles. The particles with different species are interacting via a delta-function potential. The Hamiltonian of the systems is,
\begin{align}
H_{\rm dd} 
& =
-\frac{1}{2 m_{1}} \, \partial_{x_{1}}^{2}
-\frac{1}{2 m_{2}} \, \partial_{x_{2}}^{2} 
-\frac{1}{2 m_{3}} \, \partial_{x_{3}}^{2} 
-\frac{1}{2 m_{3}} \, \partial_{x_{4}}^{2} 
\nonumber\\
&
+c_{0} \big[ \delta(x_{3}-x_{1}) + \delta(x_{3}-x_{2})
+ \delta(x_{4}-x_{1}) + \delta(x_{4}-x_{2})\big]
\,,
\label{eqn:H-BO-101}
\end{align}
where $m_{1}$, $m_{2}$, $m_{3}$, and $m_{4}$ are the masses and $x_{1}$, $x_{2}$, $x_{3}$, and $x_{4}$ are the coordinates of the spin-${\uparrow}$, spin-${\uparrow}$, spin-${\downarrow}$, and spin-${\downarrow}$ particles, respectively. Eq.~(\ref{eqn:H-BO-101}) can be rewritten as
\begin{align}
H_{\rm dd} 
& = 
-\frac{1}{2 \mu_{2}} \, \partial_{x}^{2}
-\frac{1}{2 \mu_{3}} \, \partial_{y}^{2}
-\frac{1}{2 \mu_{4}} \, \partial_{z}^{2} 
\nonumber\\
&+c_{0} \left[ \delta(y-x/2) + \delta(y+x/2))
+ \delta(z-x/2+\frac{m_{\downarrow} \, y}{2m_{\uparrow}+m_{\downarrow}}) 
              + \delta(z+x/2+\frac{m_{\downarrow} \, y}{2m_{\uparrow}+m_{\downarrow}})\right]
\,,
\label{eqn:H-BO-102}
\end{align}
where
\begin{align}
&m_{1} = m_{2} = m_{\uparrow}\,,\nonumber\\
&m_{3} = m_{4} = m_{\downarrow}\,,\nonumber\\
&\mu_{2} = m_{\uparrow}/2\,,\nonumber\\
&\mu_{3} = 2m_{\uparrow}m_{\downarrow}/(2m_{\uparrow}+m_{\downarrow})\,,\nonumber\\
&\mu_{4} = (2m_{\uparrow}+m_{\downarrow})m_{\downarrow}/(2m_{\uparrow}+2m_{\downarrow})\,,\nonumber\\
&x = x_{2}-x_{1}\,,\nonumber\\
&y = \frac{m_{1} x_{1}+m_{2} x_{2}}{m_{1}+m_{2}} -x_{3}\,,\nonumber\\
&z = \frac{m_{1} x_{1}+m_{2} x_{2}+m_{3} x_{3}}{m_{1}+m_{2}+m_{3}} -x_{4}\,.
\end{align}
The Schr\"odinger equation in the limit $m_{\uparrow} \to \infty$ is,
\begin{align}
&\left[-\frac{1}{2 \mu_{3}} \, \partial_{y}^{2}
+c_{0}\,\delta(y-x/2) + c_{0}\,\delta(y+x/2)
- u_{3}(x) 
 \right]\phi(y,z;x)
\nonumber\\
&
+\left[
-\frac{1}{2 \mu_{4}} \, \partial_{z}^{2}  
+c_{0}\,\delta(z-x/2) + c_{0}\,\delta(z+x/2)
-u_{4}(x)
\right]\phi(y,z;x)
= 0 
\,.
\label{eqn:H-BO-103}
\end{align}
where $\phi(y,z;x)$ is the solution of Eq.~(\ref{eqn:H-BO-103}) for a fixed value of $x$, and $u_{3}(x) + u_{4}(x)$ is the energy of the system of Eq.~(\ref{eqn:H-BO-103}) for a fixed value of $x$. Using boundary conditions, the continuity of the wave functions and the discontinuity of their first derivative at $y = \pm x/2$ and $z = \pm x/2$, we obtain the following solutions,
\begin{align} 
u_{3,\ell}(x)
=\frac{1}{2 \mu_{3}} 
\left[
-\beta_{3}
+
\frac{1}{x}
W((-1)^{\ell+1} x \, \beta_{3}\, e^{x \, \beta_{3}})
\right]^{2}
\,,
\label{eqn:H-BO-104}
\end{align}
\begin{align}
u_{4,\ell}(x)
=\frac{1}{2 \mu_{4}} 
\left[
-\beta_{4}
+
\frac{1}{x}
W((-1)^{\ell+1} x \, \beta_{4}\, e^{x \, \beta_{4}})
\right]^{2}
\,,
\label{eqn:H-BO-105}
\end{align}
where $\beta_{3} = c_{0}\mu_{3}$, $\beta_{4} = c_{0}\mu_{4}$, $W(r)$ is the Lambert $W$ function, and  $\ell = 0 \, (\ell = 1)$ gives the even (odd) solution.

Now, we use $u_{3,\ell}(x)$ and $u_{4,\ell}(x)$ in Eq.~(\ref{eqn:H-BO-102}),
\begin{align}
-\frac{1}{2 \mu_{2}} \, \partial_{x}^{2}\psi(x)
+u_{3,\ell}(x)\psi(x)
+u_{4,\ell}(x)\psi(x)
=
E\psi(x)
\,.
\label{eqn:H-BO-106}
\end{align}
For the dimer-dimer system, the total wavefunction of Eq.~(\ref{eqn:H-BO-102}), $\Psi(x,y,z) = \psi(x)\phi(y,z;x)$, is anti-symmetric under exchange of $x_{1} \leftrightarrow x_{2}$. Therefore, the Born-Oppenheimer solution of Eq.~(\ref{eqn:H-BO-101}) is obtained by solving the Schr\"odinger equation for $[u_{3,1}(x) + u_{4,0}(x)]/2+[u_{3,0}(x) + u_{4,1}(x)]/2$.



\end{document}